\documentclass[%
superscriptaddress,
preprint,
 amsmath,amssymb,
 aps,
]{revtex4-2}

\usepackage{graphicx}
\usepackage{dcolumn}
\usepackage{bm}

\begin{document}

\preprint{APS/123-QED}

\title{Non-trivial quantum magnetotransport oscillations\\in pure and robust topological \mbox{$\alpha$-Sn} films}

\author{Ivan Madarevic}
\thanks{These two authors contributed equally}
 \affiliation{ 
Quantum Solid State Physics, KU Leuven, Celestijnenlaan 200D, 3001 Leuven, Belgium
}
\author{Niels Claessens} 
\thanks{These two authors contributed equally}
 \affiliation{ 
Quantum Solid State Physics, KU Leuven, Celestijnenlaan 200D, 3001 Leuven, Belgium
}
\affiliation{IMEC, Kapeldreef 75, 3001 Leuven, Belgium
}

\author{Aleksandr Seliverstov}
 \affiliation{ 
Quantum Solid State Physics, KU Leuven, Celestijnenlaan 200D, 3001 Leuven, Belgium
}

\author{Chris Van Haesendonck}
\affiliation{ 
Quantum Solid State Physics, KU Leuven, Celestijnenlaan 200D, 3001 Leuven, Belgium
}

\author{Margriet J. Van Bael}
\affiliation{ 
Quantum Solid State Physics, KU Leuven, Celestijnenlaan 200D, 3001 Leuven, Belgium
}%

\date{\today}

\begin{abstract}
We report experimental evidence of topological Dirac fermion charge carriers in pure and robust \mbox{$\alpha$-Sn} films grown on InSb substrates. This evidence was acquired using standard macroscopic four-point contact resistance measurements, conducted on uncapped films with a significantly reduced bulk mobility. We analyzed and compared electrical characteristics of the constituting components of the \mbox{$\alpha$-Sn}/InSb sample, and propose a three-band drift velocity model accordingly. A surface band, with low carrier density and high mobility, is identified as the origin of the observed Shubnikov -- de~Haas oscillations. The analysis of these quantum oscillations results in a non-trivial value of the phase shift $\gamma =0$, characteristic for topologically protected Dirac fermions. For the same uncapped samples we estimate the momentum relaxation time $\tau\approx 300\ \mathrm{fs}$, which is significantly larger in comparison with the previous reports on grown \mbox{$\alpha$-Sn} films.

\end{abstract}

\maketitle

The alpha phase of Sn (\mbox{$\alpha$-Sn}) is a \textit{diamond structured} crystal which is a zero band gap semimetal. Due to its suitable band structure, \mbox{$\alpha$-Sn} was predicted as a material capable of acquiring several topological phases~\cite{Huang2017_tensileZZ, Zhang_Luttinger_Sn2018}. Crystal strain can bring about the topologically protected states in \mbox{$\alpha$-Sn}, more specifically: in-plane tensile strain turns it into a 3D topological insulator (3DTI), while in-plane compressive strain converts it into a topological Dirac semimetal (DSM). Topologically protected bands of \mbox{$\alpha$-Sn} crystalline thin films were characterized by experiments during the last decade. This was particularly done using angle-resolved photoemission spectroscopy (ARPES), which revealed a Dirac cone near the Fermi level in the surface electronic structure of Te and/or Bi doped \mbox{$\alpha$-Sn} thin films, compressively strained on InSb substrates~\cite{Barfuss2013, ohtsubo2013dirac, RogalevPhysRevB.95.161117,Scholz-PhysRevB.97.075101, Barbedienne-PhysRevB.98.195445}. Interestingly, the most recent (soft X-ray) ARPES investigations have suggested the existence of a topologically protected bulk band (DSM phase) in compressively strained \mbox{$\alpha$-Sn} films~\cite{Xu2017-alphaSn111-semi,RogalevPhysRevB.95.161117,Rogalev2019}. These reports confirm that this material is an excellent platform to explore several topological phases~\cite{Huang2017_tensileZZ, Zhang_Luttinger_Sn2018}, making it very interesting for possible spintronics~\cite{Kondou_S-C_2016, Rojas_PhysRevLett.116.096602} and quantum computing applications~\cite{KITAEV_qubit_2003,TewariSarma_qubit_PRL_2007,FuKane_proxyMajor_PRL_2008}. Moreover, as a very well structurally defined non-toxic elemental material with a robust topological phase, \mbox{$\alpha$-Sn} could be more favorable for applications when compared to prototypical DSM binary compounds Na$_3$Bi~\cite{Na3Bi_Liu2014,Na3Bi_Xu2014} and Cd$_3$As$_2$~\cite{Cd3As2_Borisenko2014, Cd3As2_Liu2014,neupane2014observation}.



There are recent reports indicating topological magnetotransport in \mbox{$\alpha$-Sn}. In the case of \mbox{$\alpha$-Sn} on InSb there is, until now, one report~\cite{Barbedienne-PhysRevB.98.195445} suggesting topologically protected carriers based on magnetotransport measurements. However, these experiments were conducted on a relatively complicated system (Bi-doped \mbox{$\alpha$-Sn$\mid$InSb$\mid$GaAs}), in which the
Dirac point was positioned below the Fermi level. There is another report on the \mbox{$\alpha$-Sn} magnetotransport properties in the case of a thin film grown on CdTe~\cite{Vail_Sn_CdTe_trans2019}. The use of insulating CdTe should allow more straightforward detection of the \mbox{$\alpha$-Sn} topological bands. However, both Cd and Te bond with Sn (by metallic and covalent bonding respectively), complicating the morphology of the films, disturbing its epitaxial growth and the introduction of strain~\cite{Tu_SdH_1989, Vail_Sn_CdTe_trans2019}. At this moment, no unambiguous experimental evidence has been presented for topological Dirac charge carriers in completely pure (no dopants) \mbox{$\alpha$-Sn} films.

Very recently we presented~\cite{madarevic2020} a straightforward method to grow pure, compressively strained \mbox{$\alpha$-Sn} films on InSb(100) substrates. The structure and morphology of these films were fully characterized and the presence of a Dirac cone was confirmed by ARPES. Moreover, these films demonstrated an excellent surface quality and a notable robustness against ambient conditions. In this letter we report prominent accessibility of the topologically protected Dirac charge carriers in pure and uncapped \mbox{30 nm} thick \mbox{$\alpha$-Sn} films compressively strained on InSb(100)~\cite{madarevic2020}. This was made possible due to the decreased mobility of the trivial bulk charge carriers in the grown films -- most likely a consequence of increased grain-boundary barrier due to the small grain size. We present evidence for the existence of the Dirac charge carriers, as an absolutely intrinsic property of the topological \mbox{$\alpha$-Sn} films. This evidence is based on our analysis of the Shubnikov -- de~Haas (SdH) oscillations originating from a surface band. The Dirac charge carriers exhibit a significantly enhanced relaxation time compared to previous reports~\cite{Barbedienne-PhysRevB.98.195445, Rojas_PhysRevLett.116.096602}.


For the unambiguous interpretation of electronic transport experiments on composite systems (i.e. film/substrate) it is generally required to analyze the transport properties of the constituting components. Therefore, along with the transport measurements on the 30~nm \mbox{$\alpha$-Sn$\mid$InSb(100)} system, we conducted the same measurements on a non-covered InSb(100) substrate, prepared according to the procedure presented in our previous report~\cite{madarevic2020}. The magnetotransport experiments were conducted at a temperature of \mbox{1.5 K}, with silver paste contacts in a four-point geometry, at constant current (between \mbox{50 and 100 $\mu$A}), and magnetic field ($B$) up to \mbox{8 T}.

\begin{figure}
\includegraphics[width=236.84843pt,keepaspectratio]{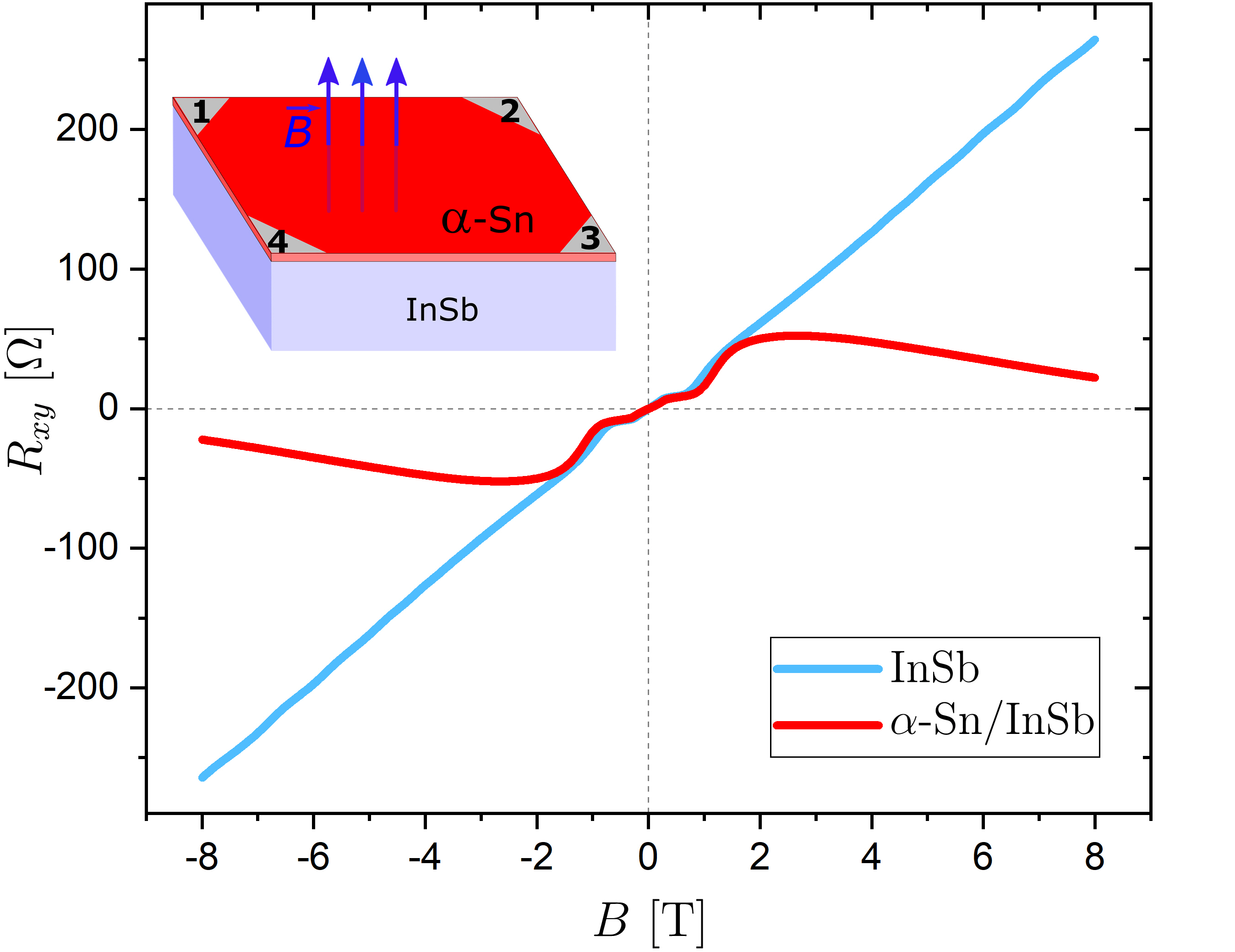}
\caption{\label{fig1} Measured off-diagonal magnetoresistance tensor element ($R_{xy}$) of a \mbox{30 nm} thick film of \mbox{$\alpha$-Sn(100)} on InSb(100), and a non-covered InSb(100) substrate. Measurements were done using the Van der Pauw technique~\cite{VdP1958method}. The inset shows the four-point contact (1~--~4) configuration on the surface of the sample.}
\end{figure}

Figure~\ref{fig1} depicts the results of the conducted Hall effect measurements, revealing a clear qualitative difference between the off-diagonal magnetoresistance ($R_{xy}$) of the InSb(100) substrate and the grown \mbox{$\alpha$-Sn$\mid$InSb(100)} sample. While the substrate’s off-diagonal magnetoresistance shows a linear behavior ($n$-type), the grown sample clearly displays a non-linear Hall effect, indicating the presence of multiple types of carriers. At low fields (\mbox{$<$ 2 T}), the substrate is shunting the \mbox{$\alpha$-Sn} film, and the behavior of $R_{xy}$ is slightly non-linear. From the results of our detailed study of the transport properties of the substrate, there are some indications that the substrate preparation (Ar$^+$ plasma etching and thermal annealing~\cite{madarevic2020}) leaves it electronically heterogenous in depth, which may cause such non-linear behavior. However, it is beyond the scope of this letter to discuss the cause of the non-linear behavior at low fields. Since it can be attributed to the substrate, for which the linear $n$-type behavior becomes dominant at higher fields, we omit the low field region from further analysis.

\begin{figure}
\includegraphics[width=232.84843pt,keepaspectratio]{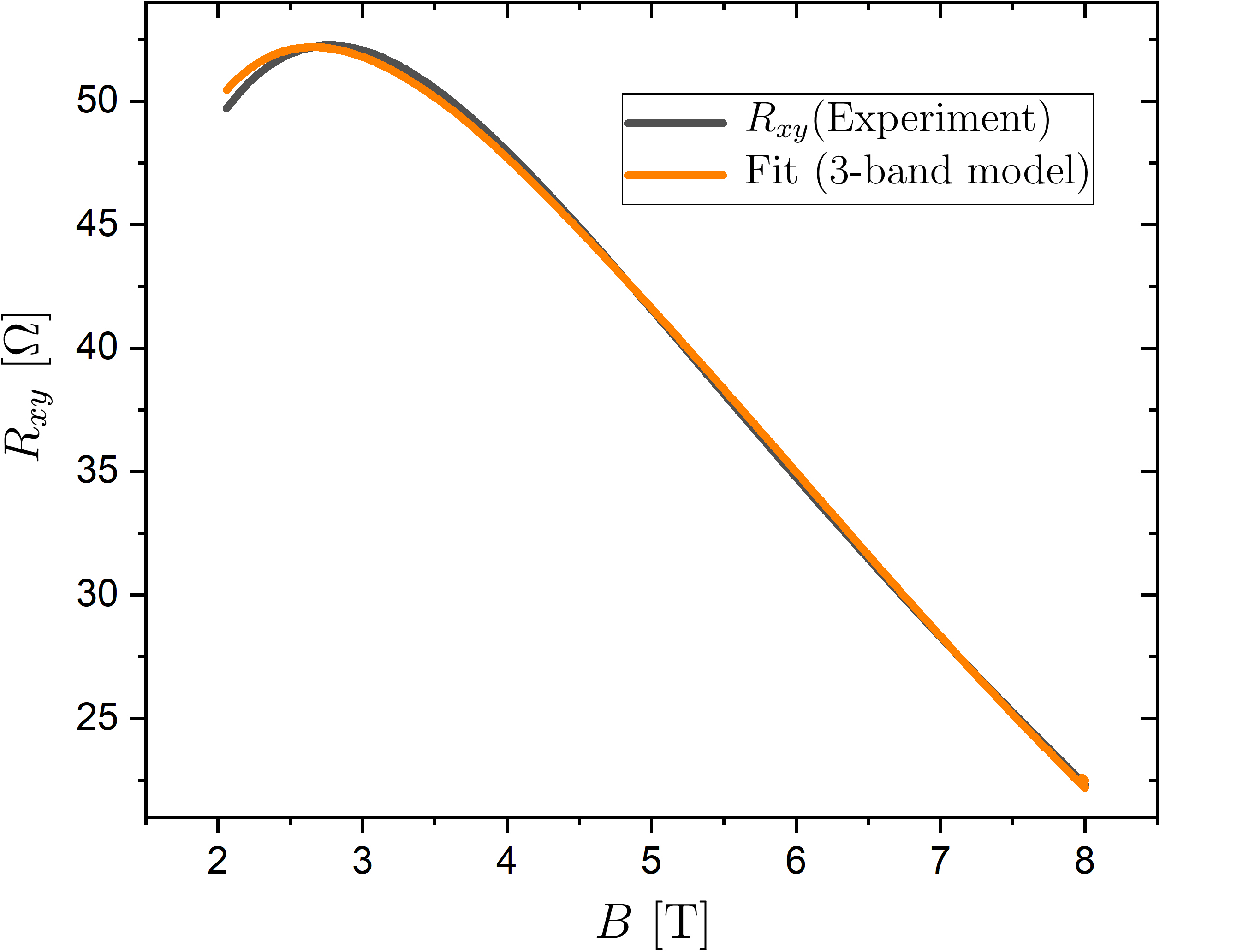}
\caption{\label{fig2} Fit based on the three-band drift velocity model (orange) for the high-field off-diagonal magnetoresistance above \mbox{2 T} (black) of a \mbox{30 nm} thick film of \mbox{$\alpha$-Sn} on InSb(100).}
\end{figure}

The method to identify different contributions leading to the observed Hall effect is to fit a multi-band drift velocity model to the measured $R_{xy}$ data. $R_{xy}$ can, in general, be expressed using the following equations:
\begin{equation}
R_{xy}=\ \frac{{\sigma }_{xy}}{{{\sigma }_{xy}}^2+{{\sigma }_{xx}}^2}~,
\end{equation}
with the conductance matrix components of the different contributions ($i$) to $\sigma$ given by:
\begin{equation}
{\sigma }^i_{xx}=\ \frac{n^i_s\cdot e \cdot{\mu }^i}{1+{\left({\mu }^i\cdot B\right)}^2}~,
\end{equation}
\begin{equation}
{\sigma }^i_{xy}=\ \pm \frac{n^i_s\cdot e\cdot {\left({\mu }^i\right)}^2\cdot B}{1+{\left({\mu }^i\cdot B\right)}^2}~,
\end{equation}
under the following constraint for the measured sheet resistance at zero field:
\begin{equation}
R_s=\ \left(\sum^k_1{n^i_s\cdot }e\cdot {\mu }^i\right)^{-1}~. {\label{eq:Rs}}
\end{equation}
Fitting the measured $R_{xy}$ behavior of the \mbox{$\alpha$-Sn} sample to a two-band ($k=2$) drift velocity model results in one $p$-type band ($i=p$) and one $n$-type band ($i=n$), with sheet carrier concentrations $n^p_s=3.2(1) \cdot{10}^{14}\ \mathrm{cm}^{-2}$, $n^n_s=1.65(4)\cdot{10}^{13}\ \mathrm{cm}^{-2}$, and mobilities ${\mu}^p=0.184(2)\cdot 10^{3}\ \mathrm{cm^2V^{-1}s^{-1}}$, ${\mu}^n=1.4(8)\cdot{10}^{4}\ \mathrm{cm^2V^{-1}s^{-1}}$. In this case, the reduced chi-square value (${\chi}^2$) for the fit equals 0.78, at a $R_{xy}$ relative error of 5~\%. On the other hand, repeating the procedure for a three-band ($k=3$) drift velocity model (Fig.~\ref{fig2}) results in two $p$-type bands ($i=p1,~p2$) and one $n$-type band ($i=n$), with sheet carrier concentrations $n^{p1}_s=3.2(1)\cdot{10}^{13}\ \mathrm{cm}^{-2}$, $n^{p2}_s=1.3(2) \cdot{10}^{12}\ \mathrm{cm}^{-2}$, $n^n_s=1.65(4)\cdot{10}^{13}\ \mathrm{cm}^{-2}$, and mobilities ${\mu}^{p1}=0.177(2)\cdot 10^{3}\ \mathrm{cm^2V^{-1}s^{-1}}$, ${\mu}^{p2}=3.47(12)\cdot 10^{3}\ \mathrm{cm^2V^{-1}s^{-1}}$, ${\mu}^n=1,4(8)\cdot{10}^{4}\ \mathrm{cm^2V^{-1}s^{-1}}$, with ${\chi}^2=0.53$ at the same relative error. Considering these values, it is tempting to choose the latter model over the first one. Indeed, knowing the film/substrate thickness~\cite{madarevic2020}, the bands with the large number of carriers ($n$ and $p1$) are typical bulk bands expected for InSb~\cite{Madelung_book} and \mbox{$\alpha$-Sn} grown on InSb~\cite{FARROW1981507}, which exhibits a reduced mobility compared to the bulk single crystal case~\cite{Ewald1958}. At the same time, the band with the high mobility and low carrier density ($p2$) may be identify as a possible topological surface band of \mbox{$\alpha$-Sn}~\cite{TIbookMolen_2013}. Although the second model fits the data better, a nicer fit on its own does not prove the correctness of the model.

\begin{figure}
\includegraphics[width=236.84843pt,keepaspectratio]{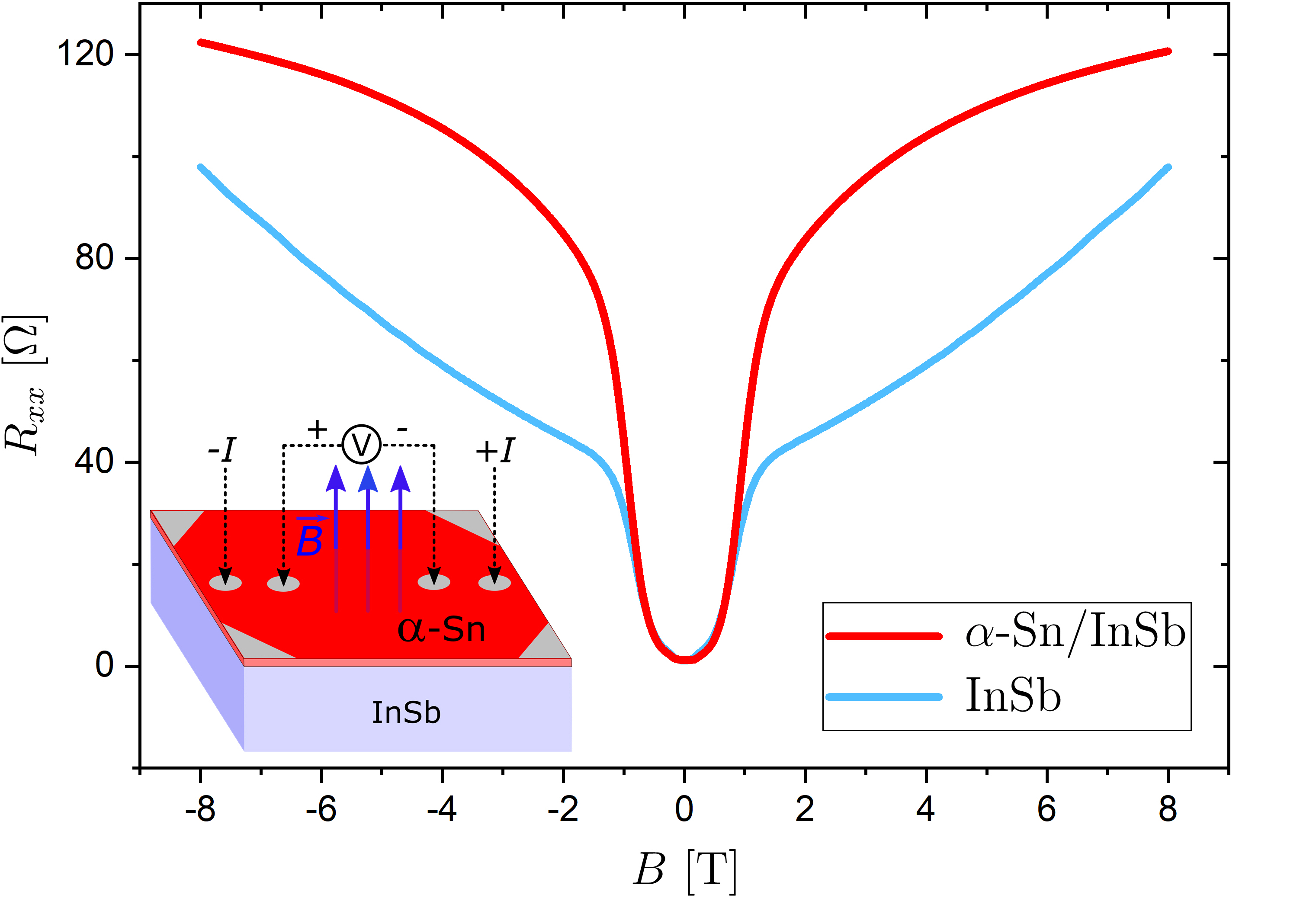}
\caption{\label{fig3} Measured diagonal magnetoresistance tensor element ($R_{xx}$) of a \mbox{30 nm} thick film of \mbox{$\alpha$-Sn(100)} on InSb(100) and a non-covered InSb(100) substrate, with the magnetic field perpendicular to the film/substrate surface. The inset shows the four-point contact configuration on the surface of the sample.}
\end{figure}

\begin{figure}
\includegraphics[width=236.84843pt,keepaspectratio]{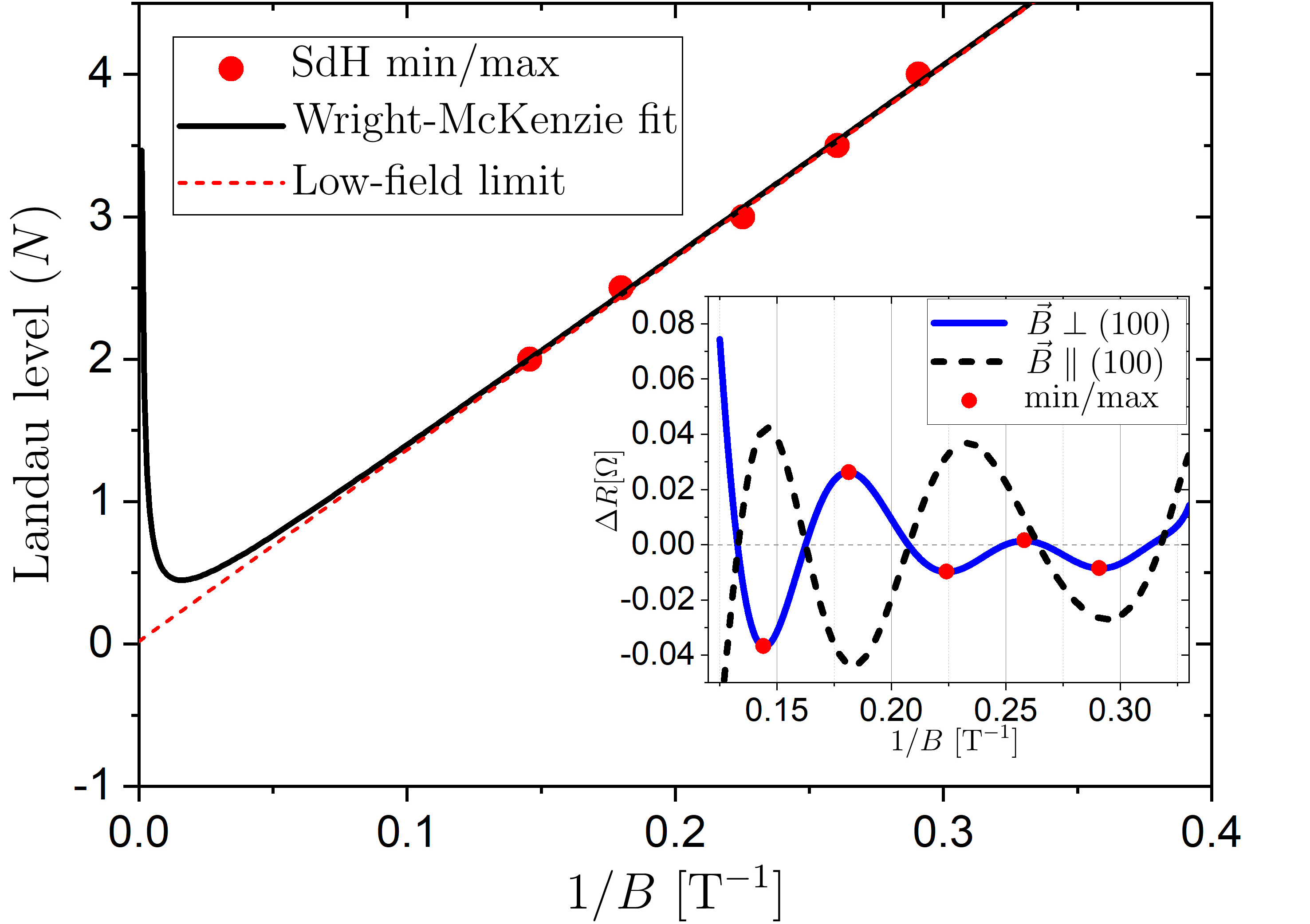}
\caption{\label{fig4} Landau index plot of the magnetoresistance oscillations after removal of the background ($\vec{B} \perp (100)$). The Landau index for the minima and maxima of the observed SdH oscillations is plotted against the inverse of the field strength (red dots) at which they are observed. The fit of Eq.~(\ref{eq:W-Mc}) (black line) through the data points is shown together with the low field limit ($A_2=0$) (red dashed line). The inlay shows magnetoresistance oscillations in the $\vec{B} \perp (100)$ (blue line) and the $\vec{B} \parallel (100)$ orientation (black dashed line).}
\end{figure}

To exclude one of the two models, we extracted the SdH oscillations from the diagonal magnetoresistance $R_{xx}$ measurement of a grown \mbox{$\alpha$-Sn} sample (Fig.~\ref{fig3}). The experiments were again conducted at a temperature of 1.5~K and with magnetic field (up to 8 T) applied perpendicularly to the surface of the samples ($\vec{B} \perp (100)$), with macroscopic silver paste contacts arranged in a linear four-point geometry. Figure~\ref{fig4} (inlay) depicts magnetoresistance oscillations ($\Delta R$) of a grown \mbox{$\alpha$-Sn} sample, extracted after removing the background which was obtained from a $3^{\mathrm{rd}}$ order polynomial fit. The maximum of the fast Fourier transform (FFT) of $\Delta R$ vs. $1/B$ provides an estimate of the characteristic frequency $f=14(6)\ \mathrm{T}$ of the oscillations. A more accurate value of 13.5(8) T was extracted from the Landau index plot (Fig.~\ref{fig4}), as shown later.

Assuming the oscillations originate from a bulk band, the estimated value for $f$ would correspond to a carrier density of $n=(4\pi f/\mathrm{\Phi _0})^{{3}/{2}}/{3{\pi }^2}=2.8(2)\cdot{10}^{17}\ \mathrm{cm}^{-3}$, where ${\mathrm{\Phi}}_0=h/e$ is the elementary flux quantum, with Planck's constant $h$ and the electron charge $e$. If we convert this to the hypothetical sheet carrier densities for such bulk bands of the \mbox{$\alpha$-Sn} film and InSb substrate, by multiplying with their thicknesses~\cite{madarevic2020}, this would give $8.4(7)\cdot {10}^{11}\ \mathrm{cm}^{-2}$ and $1.75(16)\cdot~{10}^{16}\ \mathrm{cm}^{-2}$ respectively. From these sheet carrier densities, which differ from the bulk bands extracted using the Hall effect measurements ($p1$ and $n1$, respectively), it appears that neither the two- nor three-band model would be compatible with the observed SdH oscillations if the oscillations would originate from a bulk band.

If we instead assume that the experimentally observed quantum oscillations arise from a surface band, the sheet carrier density corresponding to the observed SdH frequency is given by $n_s=2\cdot n = 2\cdot (2f/{\mathrm{\Phi}}_0) = {1.30(7)\cdot {10}^{12}\ \mathrm{cm}^{-2}}$. This carrier density value is in very good agreement with the second $p$-type band ($p2$) in the above fitted three-band drift velocity model, but in contradiction with the two-band model. Note as well that the SdH oscillations are only observable when their cyclotron frequency is larger than the inverse of the relaxation time~\cite{ashcroft1976solid} -- equivalent to $\mu >{1}/{B}\approx 3.33 \cdot 10^{3}\ \mathrm{cm^2V^{-1}s^{-1}}$ for oscillations starting at \mbox{$\sim3$~T}. This condition is satisfied for the $p2$ band in the three-band drift velocity model, but it is not met for the two-band model. Moreover, the reduced mobility of the trivial bulk carriers (increased grain boundary scattering) in this model (${\mu}^{p1}$) is consistent with the fact that the grown \mbox{$\alpha$-Sn} films have a granular morphology, with grain size around 10~--~20~nm~\cite{madarevic2020}. Therefore, it can be concluded that the two-band drift velocity model fails, while the three-band model explains our observations. The second $p$-type band of the three-band model ($p2$) can thus be identified as a surface band of the \mbox{$\alpha$-Sn}(100) film. The first $p$-type band ($p1$) we assign to the bulk \mbox{$\alpha$-Sn} and the $n$-type band to the bulk of the InSb substrate.

Using Eq.~(\ref{eq:Rs}) we can now estimate the sheet resistance contribution of the bands separately: $R^{~p1}_s=1.10(4)\ \mathrm{k\Omega/\Box}$ (\mbox{$\alpha$-Sn} bulk band), $R^{~p2}_s=1.4(2)\ \mathrm{k\Omega/\Box}$ (\mbox{$\alpha$-Sn} surface band) and $R^{~n}_s=27(15)\ \mathrm{\Omega/\Box}$ (InSb bulk band).

We will now investigate the possible Dirac nature of the carriers in the second $p$ band ($i=p2$). The phase offset $\gamma$ of the SdH oscillations contains the information on the dispersion of the corresponding charge carriers~\cite{shoenberg2009magnetic}. The SdH oscillations can be expressed as:
\begin{equation}
\Delta R\propto {\mathrm{cos} \left[ 2\pi \cdot \left( \frac{f}{B}-\gamma \right) \right]}~,
\end{equation}
where $\gamma$ can be related to Berry's phase acquired by the carriers within a cyclotron orbit~\cite{Fuchs_gamma_berry2010}. When the zero-field electronic dispersion would be followed by the carriers, then $\gamma =0$ (non-trivial Berry's phase of $\pi$) for Dirac fermions, while $\gamma =1/2$ (trivial Berry's phase of 0) for ``normal'' fermions~\cite{WrightMcKenzie}. The intercept of a linear fit of the Landau level index N with respect to $1/B$ then gives the $\gamma $ value. However, for finite field strengths, as pointed out by Wright and McKenzie, the fermions can no longer be expected to follow the zero-field dispersion, and a deviation from a perfect quantization, where $\gamma =\gamma (B)$, is to be expected~\cite{WrightMcKenzie}. For that reason, a more accurate low-field fit procedure was proposed to extract the zero-field phase offset:
\begin{equation}
N=\frac{f}{B}+A_1+A_2\cdot B~, {\label{eq:W-Mc}}
\end{equation}
where $f$ (characteristic frequency), $A_1$ and $A_2$ are the fit-parameters. The zero-field limit of the Eq.~(\ref{eq:W-Mc}) is equivalent with $\gamma (0)=A_1$. Therefore, the parameter $A_1$ is the quantized offset related to the zero-field Berry's phase. Figure~\ref{fig4} presents the fit of Eq.~(\ref{eq:W-Mc}) to the Landau indices of the observed SdH oscillations minima and maxima extracted from the magnetoresistance measurement with the magnetic field applied perpendicular to the sample surface plane ($\vec{B}\perp \mathrm{(100)}$). Here, due to the dominance of bulk contributions to the conductance~\cite{Xiong_LLminmax}, the minima of the SdH oscillations correspond to integer Landau level indices ($N$), while the maxima correspond to half-integer Landau level indices ($N+1/2$). The resulting fit-parameters are $A_1=0.02(15)$ and  $A_2=0.00(6)$, and $f=13.5(8)\ \mathrm{T}$. Contrary to the near-zero $A_1$ value for the \mbox{$\alpha$-Sn} samples, the fit of the SdH oscillations extracted from the magnetoresistance measurement on a non-covered InSb substrate gave $A_1=0.52(13)$, $A_2=-0.04(2)$ and $f=39.4(8)\ \mathrm{T}$. The characteristic frequencies were determined by a linear fit of the Landau index plot (Fig.~\ref{fig4}) and held fixed as the linear fit yields more accurate results on the frequency compared to the estimation of the FFT maximum of $\Delta R$, which is the usual procedure in order to reduce the number of fit parameters. As the analysis in the case of \mbox{$\alpha$-Sn} gives $A_1\approx0$, it can be concluded that the second $p$-type band ($i=p2$) indeed has a topological Dirac fermion nature.
Based on this conclusion, we can now estimate the value of the characteristic Fermi wavevector for this $p2$ band as $k_F=\sqrt{{4\pi}{f/\Phi_0}}=0.203(6)\ \mathrm{nm}^{-1}$. Assuming a linear energy dispersion~\cite{neto2009electronic} and using the previously extracted Fermi velocity value~\cite{madarevic2020}, we then estimate the Fermi energy at $E_F=\hslash k_F v_F=67(2)\ \mathrm{meV}$. These values, estimated from the transport measurements, are in agreement with the ones extracted from our recent ARPES and scanning tunneling spectroscopy (STS) study ($k_F ^{~ARPES}\sim0.18\ \mathrm{nm}^{-1}$, $E_F ^{~ARPES}\sim 60\ \mathrm{meV}$ and $E_F^{~STS}\sim 70\ \mathrm{meV}$ below the Dirac point)~\cite{madarevic2020}.
%

From the calculated $R^{~p2}_s$ value, using the resistance expression for non-degenerate 2D bands with a linear dispersion, i.e., $R^{p2}_s=(4\hbar ^2 \pi)/(e^2 E_{F} \tau)$~\cite{gantmakher2012carrier},
we estimate the momentum relaxation time $\tau = 300(40)\ \mathrm{fs}$. This value is about five times larger than the one estimated for \mbox{$\alpha$-Sn} films with an AlO$_x$ capping layer~\cite{Barbedienne-PhysRevB.98.195445}, and almost two orders of magnitude larger compared with \mbox{$\alpha$-Sn} films covered with Ag~\cite{Rojas_PhysRevLett.116.096602}, indicating that there are fewer parallel momentum relaxation channels in our samples.

 In the case of a 2D gas of Dirac fermions existing on top (and at the bottom) of the grown \mbox{$\alpha$-Sn} film, one expects the SdH oscillations to vanish for the $\vec{B} \parallel \mathrm{(100)}$ alignment. However, for our samples this is not the case. Surprisingly, in the same type of grown \mbox{$\alpha$-Sn} samples, SdH oscillations were also observed (inlay of Fig.~\ref{fig4}), when the magnetic field is applied parallel to the film surface (equivalent to $\vec{B} \perp \mathrm{(010)}$). These SdH oscillations have a characteristic frequency $f=10.1(15)\ \mathrm{T}$, which is quite similar to the $\vec{B} \perp \mathrm{(100)}$ case, and which cannot be attributed to InSb or \mbox{$\alpha$-Sn} bulk bands.
The phase offset analysis of these oscillations gave a value $A_1=0.1(3)$ close to the non-trivial Berry's phase of $\pi$. From the topographic details of our samples~\cite{madarevic2020} we know that the surface of the film is not atomically flat, but in fact consists of crystalline grains with non-zero roughness. Therefore, a significant portion of the surface area is never in parallel alignment with the applied magnetic field.

 

 


All of the above presented results do strongly support the existence of topological Dirac fermion charge carriers, but our discussion so far has not yet specified which type of Dirac material, 3DTI or DSM, our \mbox{$\alpha$-Sn} films belong to. In the case of the 3DTI \mbox{$\alpha$-Sn}, a surface phase with a clear topological Dirac signature should arise in the electrical transport experiments~\cite{Huang2017_tensileZZ,Zhang_Luttinger_Sn2018}. The detected $p2$ surface band appears to agree with the 3DTI picture. However, this phase is predicted to emerge for tensile strained \mbox{$\alpha$-Sn}, while our samples are compressively strained~\cite{madarevic2020}. On the other hand, the most recent ARPES studies confirmed that for in-plane compressive strain (which is the case for our samples) \mbox{$\alpha$-Sn} is a DSM~\cite{Xu2017-alphaSn111-semi, Rogalev2019}. This portrays \mbox{$\alpha$-Sn} as a material which is topologically non-trivial regardless of the type of strain. For in-plane compressively strained film, the \mbox{$\alpha$-Sn} DSM phase should host two Dirac nodes, projected to the (100) plane as a single Dirac cone~\cite{Huang2017_tensileZZ}. Topological surface states and Fermi arcs (between multiple Dirac points) can form as a consequence of strain, and/or the applied magnetic field~\cite{Zhang_Luttinger_Sn2018}. 
However, magnetotransport experiments, aside of the detected topological Dirac-like 2D surface band, do not show the existence of the (DSM) Dirac bulk charge carriers in our samples (e.g.~through an additional frequency in the SdH oscillations). For our experiments the lack of sensitivity to the topology of the bulk is due to the low mobility of the film bulk charge carriers and the obvious low resistance of the InSb substrate (resulting in pronounced shunting of the \mbox{$\alpha$-Sn} film). 
For the above stated reasons, based on our transport results it is not possible to unambiguously resolve the type of topological phase (3DTI or DSM) responsible for the Dirac cone observed in our \mbox{$\alpha$-Sn} films~\cite{madarevic2020}.

We conclude that we obtained evidence for the existence of topological Dirac fermion charge carriers in uncapped pure and robust \mbox{$\alpha$-Sn}(100) films. A $p$-type surface band with low carrier density and high mobility has been identified to be the origin of the observed SdH oscillations. To analyze these quantum oscillations, Landau index plots were fitted, revealing a non-trivial value of the phase shift, characteristic for topologically protected Dirac fermions. Such findings strongly support earlier theoretical reports of topologically protected charge carriers being an intrinsic property of strained \mbox{$\alpha$-Sn}. A~significantly longer momentum relaxation time of the detected carriers suggests more prominent accessibility of the topological properties in \mbox{$\alpha$-Sn(100)} grown on InSb(100). The remarkable fact that it is possible to detect topological Dirac fermion-like transport in our samples, using standard macroscopic four-point resistance measurements, provides unique opportunity to further investigate and apply this elemental topological material. 

We thank J. Van de Vondel, B. Raes and G. Lippertz for valuable discussions. This work was supported by the Research Foundation -- Flanders (FWO) and by the KU Leuven C1 program grants No. C14/18/074 and  C12/18/006.



\bibliography{apssamp}

\end{document}